\documentclass[11pt,a4paper]{article}
\pdfoutput=1
\usepackage{jcappub}
\usepackage{slashed}
\usepackage{stackengine}
\def\be{\begin{equation}}
\def\ee{\end{equation}}
\def\bea{\begin{eqnarray}}
\def\eea{\end{eqnarray}}

\begin{document}

\title{Entropic Mechanics: towards a stochastic description of quantum mechanics } 

\author{Vitaly Vanchurin}

\emailAdd{vvanchur@d.umn.edu}

\date{\today}

\affiliation{Department of Physics, University of Minnesota, Duluth, Minnesota, 55812 \\
Duluth Institute for Advanced Study, Duluth, Minnesota, 55804}

\abstract{We consider a stochastic process which is (a) described by a continuous-time Markov chain on only short time-scales and (b) constrained to conserve a number of hidden quantities on long time-scales. We assume that the  transition matrix of the Markov chain is given and the conserved quantities are known to exist, but not explicitly given. To study the stochastic dynamics we propose to use the {\it principle of stationary entropy production.} Then the problem can be transformed into a variational problem for a suitably defined ``action'' and with time-dependent Lagrange multipliers. We show that the stochastic dynamics can be described by a Schr\"odinger equation, with Lagrange multipliers playing the role of phases, whenever (a) the transition matrix is symmetric or the detailed balance condition is satisfied, (b) the system is not too far from the equilibrium and (c) the number of the conserved quantities is large. }

\maketitle

\section{Introduction}

From the early days of quantum mechanics physicists tried to come up with a classical or statistical model which would explain the bizarre prediction of quantum mechanics. The literature on the subject is rather vast and we are not going to discuss it here. The interested reader is referred to classic works \cite{Fenyes, Nelson} as well as to more recent books on the subject \cite{Adler, Caticha, tHooft} (and references therein) where some new and original proposals are also presented (e.g. trace dynamics \cite{Adler}, entropic dynamics \cite{Caticha}, cellular automata interpretation \cite{tHooft}). Perhaps it is worth mentioning that any attempts to derive quantum mechanics from classical or statistical mechanics should be taken with great caution due to severe experimental constraints on local hidden variables theories \cite{Bell1, Bell2} imposed by Bell's inequalities \cite{Bell}. We are not going to discuss Bell's inequalities here either just because we are not yet at the stage of matching experimental results. See, however Ref. \cite{Duality}, for a recently proposed duality between a quantum system of spinors and a classical system of scalars.  The coupling between scalars is non-trivial (e.g. a model on $2$-sphere configuration space), but the locality structure of the dual systems is preserved and so the duality can potentially be used to study the Bell's inequalities in context of local hidden variables theories.

In this paper, our main goal is to identify the conditions under which a statistical system would evolve according to rules of quantum mechanics. More precisely, we want to construct a stochastic process (which need not be Markovian) whose dynamics would be (if not exactly, but approximately) described by a Schr\"odinger equation. The stochastic process will be assumed to obey Markovian dynamics on the shortest time-scales (e.g. Planck time) and at the same time the dynamics will be constrained to conserve a large number of conserved quantities. As a result, the overall process may no longer be Markovian, but is a martingale in a sense that expectation values of the conserved quantities in the initial state remain unchanged throughout evolution. Whether such a process would be a generic consequence of coarse-graining  (or of lumping of states) is an important question which deserves a separate study. We will provide an example of a physical system for which these two conditions (i.e. Markovian on short time-scales and martingale on long time-scales) are satisfied, but for the most part of the paper we shall assume that such a process exists and the main problem will be to determine the most probable path that a statistical state of such a system would take. 

To tackle the problem we propose to use the {\it principle of stationary entropy production}. We are not going to prove the principle but it is, once again, something which deserves a separate analysis. The intuition behind the principle is that (a) at late times the most stable periodic orbits (or trajectories) are the ones for which the entropy production is smallest, but (b) at early times the trajectories with the largest entropy production are the ones that are more likely to find the periodic orbits faster. Fortunately,  when it comes to the equation of motion derived in the paper (i.e. an approximate Schr\"odinger equation) it does not matter whether the entropy production is maximized or minimized. Note, however, the apparent similarities between the stationary action principle (also known as least action principle) in classical mechanics and the stationary entropy production principle which we propose to use in context of statistical systems, or what we shall call {\it entropic mechanics}. In classical mechanics, action is extremized along classical paths and in entropic mechanics, entropy production is what must be extremized along statistical paths. In fact, our framework of entropic mechanics is very similar in spirit to both classical mechanics and also entropic dynamics \cite{Caticha}, but the underlying principles (i.e. stationary entropy production principle  vs. stationary action principle or maximum entropy principle) are not the same. Of course, it is possible that all these principles are intimately related to each other. 

The paper is organized as follows. In the next section we define a stochastic process which is described by a Markov process on only short time-scales and highly constrained on longer time-scales. To study the process we propose the stationary entropy production principle which is introduced in Sec. \ref{sec:principle}. In Sec. \ref{sec:production} we apply the condition of detailed balance and derive an expression for entropy production near equilibrium. In Sec. \ref{sec:wavefunction} we introduce an auxiliary wave function and in Sec. \ref{sec:Schrodinger} we derive an equation which governs its dynamics in the limit of a large number of constraints. In Sec. \ref{sec:conclusion} we discuss the main results of the paper.

\section{The Stochastic Process}\label{sec:process}

Consider a continuous-time Markov chain described by a master equation
\be
\frac{d p_m}{dt} = \sum_{n} \Delta_{mn} p_n \label{eq:master},
\ee
where $p_n$  is the probability to find the system in state $n$ and $\Delta_{mn}$ is the transition rate from state $n$ to state $m$. We assume that the total number of states, $N$, is finite and summation over all states is implied unless stated otherwise (as, for example, in equation \eqref{eq:norm_condition2}). From conservation of probabilities it follows that
\be
0=\sum_m \frac{d p_m}{dt} = \sum_{m,n} \Delta_{mn} p_n,
\ee
but since this must be true for any vector $p$, including $p_n=\delta_{nl}$, we get
\be
\sum_{m} \Delta_{ml} = 0.\label{eq:norm_condition}
\ee
or 
\be
\Delta_{ll} = - \sum_{m\neq l} \Delta_{ml} \label{eq:norm_condition2}
\ee
for any $l$.  The master equation \eqref{eq:master} can be integrated to obtain a time-dependent solution $p_m(t)$. At very large times the solution will be dominated by an eigenvector of $\Delta_{mn}$ with eigenvalue zero. For irreducible Markov processes (an assumption we are going to make) the eigenvector is unique due to  Perron-Frobenius theorem.

\begin{figure}[]
\begin{center}
\includegraphics[width=0.6\textwidth]{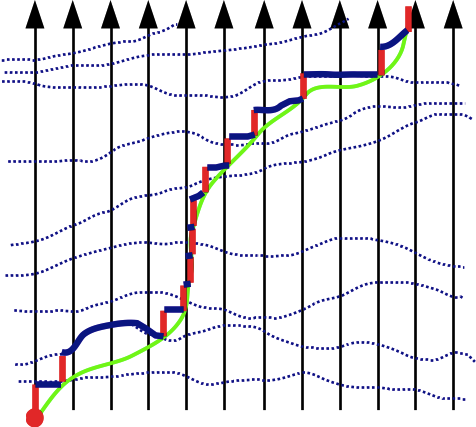}
\caption{Probability space of a stochastic process with Markovian dynamics (red vertical segments) and projections to the constrained surface (blue segments). Equi-entropic surfaces are plotted with blue dotted lines and the constrained surface with green solid line.} \label{fig:process}
\end{center}
\end{figure}

Now consider another stochastic process which is described well by the master equation \eqref{eq:master} only on short times-scales (e.g. Planck time), but for longer time-scales all that we know is that there are certain conserved quantities due to some (perhaps hidden microscopic) symmetries. We do not know what these conserved quantities are, but we know they exist. It may be useful to keep in mind a concrete (but toy-) model of a process of this type. For instance, per unit time (e.g. Planck time) the process might be such that the state first changes according to \eqref{eq:master} and then it is projected back to the surface described by $K<N$ constraints. Note that the constraint surface is $N-K$ dimensional and the equi-entropic surface  (surface on which entropy is constant) is $N-1$ dimensional and so they would generically intersect along $N-K-1$ dimensional surface. On Fig. \ref{fig:process} we provide an illustration of such a process with $N=2$, $K=1$ and $N-1-K=0$  and so there is a unique choice for the path.  Vertical lines represent the short time-scales Markovian evolution, the dotted blue lines represent equi-entropic surfaces, the thick solid green line represents a single constraint surface and the zigzagged line with alternating red and blue segments represents respectively the Markovian dynamics and projections to the constraint surface. If the projection is only a small correction to the path then the short-time dynamics would be approximated by \eqref{eq:master}, but the constraints would be satisfied on long time-scales. In general $N-1-K>0$ and so an additional principle must be imposed to single out a unique path. We will discuss one such principle in the next section.

\begin{figure}[]
\begin{center}
\includegraphics[width=0.6\textwidth]{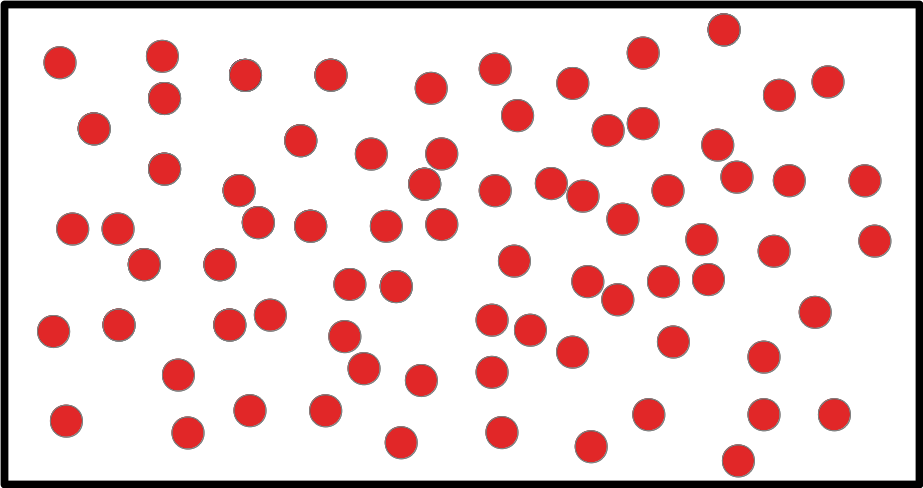}
\caption{Gas of molecules. } \label{fig:gas}
\end{center}
\end{figure}

\begin{figure}[]
\begin{center}
\includegraphics[width=0.6\textwidth]{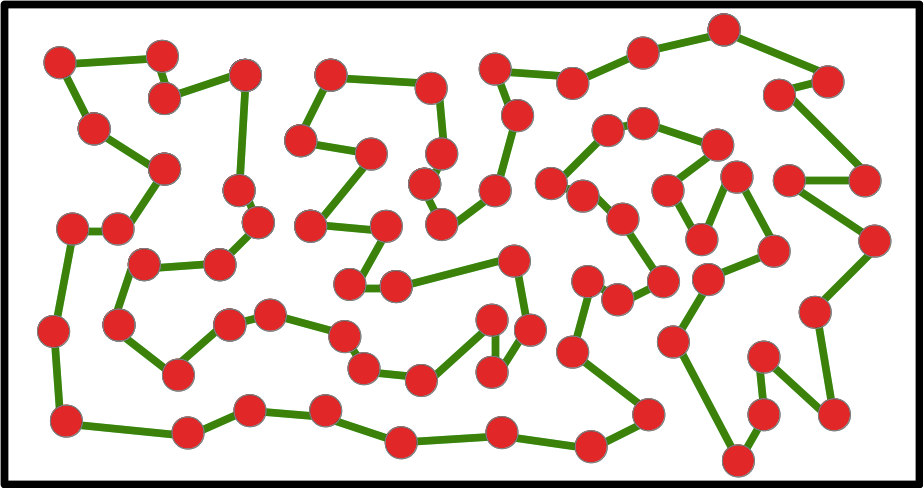}
\caption{Gas of molecules with constraints.} \label{fig:gas2}
\end{center}
\end{figure}

Perhaps a more physically-relevant example of the stochastic process can be described starting with a gas of molecules (see Fig. \ref{fig:gas}) and then introducing pairwise interactions so that the molecules form long polymer-like chains (see Fig. \ref{fig:gas2}). We are not interested in following individual molecules exactly, but only in a coarse-grained dynamics of a probability distribution of molecules. For the gas of molecules there is really only very few constraint (conservation of energy and normalization condition), but with molecules forming long chains (which we assume, for simplicity, cannot break) the overall distribution becomes highly constrained. The individual molecules can only experience an approximately Brownian motion (which is Markovian) on short time-scales as they must also respect the motion of their immediate neighbors on the chain. And the constraints imposed by the neighbors is what introduces a large number of conserved quantities on long time-scales. The situation is very similar to a fluid of strings where, in addition to the conservations of energy and momentum, an anti-symmetric tensor, which describes tangent vectors of strings, must also be conserved \cite{Vanchurin2, Schubring, Schubring2}.

In what follows, we are not going to consider a particular process, but a class of processes for which the short-time dynamics is described by \eqref{eq:master}  and the long-time dynamics is constrained to conserve a large number of conserved quantities. The conserved quantities will be denoted by $\Theta^{(\alpha)}_n$'s and their expectation values as
\be
\theta_\alpha = \sum_{n} p_n \Theta^{(\alpha)}_n \label{eq:constraints}
\ee
for $\alpha=1,...,K $. Note that the normalization condition can also be imposed as a conserved quantity with, for example, $\Theta^{(1)}_n =1$ and expectation value 
\be
\theta_1  = \sum_{n} p_n\Theta^{(1)}_n = \sum_{n} p_n  = 1.\label{eq:normalization2}
\ee
As in the case of a pure Markov process, we are interested in finding $p_n(t)$ for any $t$, but now the problem does not have a unique solution and some principle must be postulated in order to single out a unique (time-dependent) distribution $p_n(t)$.

\section{The Principle of Stationary Entropy Production}\label{sec:principle}

The principle of maximum entropy \cite{Jaynes}  states that when a number of different probability distributions are consistent with the same set of constraints, the most reasonable choice corresponds to a distribution which has the largest Shannon entropy 
\be
S \equiv -  \sum_{n}p_n \log p_n.
\ee 
Then the problem can be solved using the method of Lagrange multipliers. If we define a ``Lagrangian'' 
\be
{\cal L}(\lambda_1, ...., \lambda_K; p_1, ..., p_N)  \equiv-  \sum_{n}p_n \log p_n+  \sum_\alpha \lambda_\alpha \left ( \sum_{n} p_n \Theta^{(\alpha)}_n  - \theta_\alpha  \right ),\label{eq:Lagrangian}
\ee
then at a (local) maxima of $\cal L$ the partial derivatives with respect to $p_n$'s and $\lambda_\alpha$'s must vanish
\bea
0 &=& \frac{\partial {\cal L}}{\partial p_n} =   \sum_\alpha \lambda_\alpha \Theta^{(\alpha)}_n  - \log p_n- 1 \\
0 &=&  \frac{\partial {\cal L}}{ \partial \lambda_\alpha} = \sum_{n} \Theta^{(\alpha)}_n p_n - \theta_\alpha\label{eq:constraints2}
\eea
and, thus, the maximum entropy distribution would be given by 
\be
p_n  = \exp\left(-1+ \sum_\alpha \lambda_\alpha \Theta^{(\alpha)}_n \right) \label{eq:solution}
\ee
with  Lagrange multipliers  $\lambda_{\alpha}$ determined from the constraints \eqref{eq:constraints2}.  Using this principle a number of fundamental results of statistical mechanics can be derived from information theory, but it remains unclear what role the principle may play in context of quantum mechanics. (See, however, Ref. \cite{Caticha} where the principle was used to study a possible emergence of quantum mechanics from information theory.).

In our problem we are not interested in time-independent equilibrium states (and not even in time-independent non-equilibrium steady states), but in how the state evolves in time. In other words, we must obtain a solution for the entire path $p_m(t)$ from some initial time $t=0$ to some final time $t=T$ and thus we are forced to use another principle. In this paper we shall consider the following principle: \\

{\bf Principle of Stationary Entropy Production}: {\it The path taken by the system is the one for which the entropy production is stationary. }\\
\\
The principle can be thought of as a generalization of both, the maximum entropy principle \cite{Jaynes} and the minimum entropy production principle \cite{Prigogine,Klein} (which is often used to study steady states in non-equilibrium thermodynamics.) The main difference, however, is that instead of specifying only a single state (e.g. an equilibrium or a steady state) the principle of stationary entropy production is supposed to describes the entire path $p_m(t)$. 

According to the second law of thermodynamics, the entropy must grow and all that the principle says is that this growth has to be extremized (either as slow as possible or as fast as possible). In other words what we want to extremize is the entropy production or the total entropy change, 
\bea
\Delta S & \equiv&  S(T) - S(0) \notag\\
 &=& \int_0^T dt \frac{d S(t)  }{ dt} \notag\\
&=&  - \int_0^T dt \sum_{m}\frac{d p_m(t)  }{dt} \left ( \log p_m(t)    +1 \right ) \notag\\
&=&  - \int_0^T dt \sum_{m,n} p_n(t)  \Delta_{mn} \log p_m(t), \label{eq:entropy_production}
\eea
subject to whatever constraints. Note that we have explicitly assumed that the entropy production is due entirely to our Markovian dynamics on the short time-scales and contributions coming from restricting trajectories to remain on the constraint surface are negligible (see Fig \ref{fig:process}). More generally, the entropy production may also include a term which describes an outgoing flux of entropy, but for the case of a symmetric transition matrix \eqref{eq:symmetry}, considered in this paper, the total flux would be zero and then equation \eqref{eq:entropy_production} would still describe the total entropy production.

Given \eqref{eq:entropy_production} we want to apply the principle of stationary entropy production to find a path which corresponds to either the least or the greatest amount of produced entropy. To accomplish this task, we define not a ``Lagrangian'', but an ``action'' that is extremized along the paths of stationary entropy production,
\bea
{\cal S} &\equiv&  \int_0^T  dt\;\left ( \frac{d S}{ dt} +  \sum_\alpha \lambda_\alpha(t) \left ( \sum_{n}\; p_n(t) \Theta^{(\alpha)}_n- \theta_\alpha  \right )\right )\notag \\
&=&  \int_0^T  dt\;\left ( - \sum_{m,n} p_n(t)  \Delta_{mn} \log p_m(t)  +  \sum_\alpha \lambda_\alpha(t) \left ( \sum_{n}\; p_n(t) \Theta^{(\alpha)}_n- \theta_\alpha  \right )\right )\label{eq:action1}.
\eea
To guarantee that the constraints  \eqref{eq:constraints} are imposed at all times the Lagrange multipliers $\lambda_\alpha(t)$ are now time-dependent.

\section{Entropy Production}\label{sec:production}

The entropy production  term \eqref{eq:entropy_production} can be re-expressed as
\bea
\frac{d S}{ dt}   &=&  - \sum_{m,n}p_n \Delta_{mn} \log p_m  \notag\\
& = & - \sum_m \left ( \sum_{n \neq m}p_n \Delta_{mn}  \log p_m +  p_m \Delta_{mm}  \log p_m \right ) \notag\\
& = & - \sum_m \left ( \sum_{n \neq m}p_n \Delta_{mn}  \log p_m - \sum_{n\neq m}  p_m  \Delta_{nm}    \log p_m \right ) \notag\\
& = & - \sum_m \left ( \sum_{n \neq m}p_n \Delta_{mn}  \log p_m - \sum_{n\neq m}  p_n  \Delta_{mn}    \log p_n \right ) \notag\\
& = &  \sum_m \sum_{n \neq m}p_n \Delta_{mn}  \left (\log p_n -  \log p_m \right )  \notag\\
& = &   \sum_{m,n}p_n \Delta_{mn}  \left (\log p_n -  \log p_m \right )  \label{eq:entropy_production2}.
\eea
where we used \eqref{eq:norm_condition2}. If the detailed balance condition is satisfied, i.e.
\be
\Delta_{mn} \pi_n = \Delta_{nm} \pi_m
\ee
where $\pi_m$ is the equilibrium state, then the system can be transformed into a system with symmetric transition matrix after appropriate splitting of states into substates. To simplify the analysis we will assume that the transition matrix is symmetric
\be
\Delta_{mn} = \Delta_{nm}\label{eq:symmetry},
\ee
but our analysis and conclusions will be equally valid for systems with detailed balance given that the state space was appropriately redefined. Using \eqref{eq:symmetry} the entropy production \eqref{eq:entropy_production2} becomes
\bea
\frac{d S}{ dt}   &=&  \frac{1}{2} \sum_{m,n} p_n \Delta_{mn}  \left (\log p_n -  \log p_m \right ) + \frac{1}{2} \sum_{m,n} p_n \Delta_{nm}  \left (\log p_n -  \log p_m \right )  \notag\\
&=&  \frac{1}{2}\sum_{m,n}p_n \Delta_{mn}  \left (\log p_n -  \log p_m \right ) + \frac{1}{2} \sum_{m,n}p_m \Delta_{mn}  \left (\log p_m -  \log p_n \right )  \notag\\
& = & \frac{1}{2} \sum_{m,n} (p_n - p_m ) \Delta_{mn}  \left (\log p_n  -   \log p_m  \right )  \label{eq:entropy_production3}.
\eea
Since  $\log(p)$ is a monotonically increasing function, $(p_n -p_m) (\log p_n  - \log p_m) \ge 0$, and off-diagonal elements of $\Delta_{mn} $ are non-negative, the entropy production must also be non-negative 
\be
\frac{d S}{dt} \ge 0.
\ee
This means that the second law of thermodynamics is satisfied as it should.

For a symmetric transition matrix  \eqref{eq:symmetry} the conditions \eqref{eq:norm_condition} becomes
\be
\sum_{m} \Delta_{ml} = \sum_{m} \Delta_{lm} = 0 \label{eq:symmetry2}
\ee
and, therefore, the uniform distribution $\pi_m=\frac{1}{N}$, is also an equilibrium distribution 
\be
\frac{d \pi_l}{dt} = \sum_{m} \Delta_{lm} \pi_m = \sum_{m} \Delta_{lm} \frac{1}{N} = 0.
\ee
Then near equilibrium we should be able to expand  $\log p_n$ around $p_m$, 
\be
\log(p_n) = \log(p_m + (p_n-p_m)) =  \log(p_m)+\frac{(p_n-p_m)}{p_m} -  \frac{(p_n-p_m)^2}{2 p^2_m} +  {\cal O}((p_n-p_m)^3)
\ee
which can be substituted back to the entropy production \eqref{eq:entropy_production3},
\be
\frac{d S}{ dt} = \frac{1}{2}  \sum_{m,n} \Delta_{mn} \left (\frac{(p_n-p_m)^2}{p_m} -  \frac{(p_n-p_m)^3}{2 p^2_m} + {\cal O}((p_n-p_m)^4) \right ) \label{eq:Laplacian_entropy2}.
\ee
Note, however, that 
\be
\sqrt{p_n} = \sqrt{p_m + (p_n-p_m)} = \sqrt{p_m} + \frac{ (p_n-p_m)}{2 \sqrt{p_m}} - \frac{ (p_n-p_m)^2}{8 \sqrt{p_m}^3} +    {\cal O}((p_n-p_m)^3) 
\ee
and so
\be
4 \left ( \sqrt{p_m}- \sqrt{p_n} \right )^2 = \frac{(p_n-p_m)^2}{p_m} -  \frac{(p_n-p_m)^3}{2 p^2_m} + {\cal O}((p_n-p_m)^4).
\ee
Therefore, up to the fourth (!) order in $p_n-p_m$, the entropy production is simply
\bea
\frac{d S}{ dt} &=& 2  \sum_{m, n}  \Delta_{mn} \left ( \sqrt{p_m}- \sqrt{p_n} \right )^2 \notag \\
&=&  2 \sum_{m,n}   \Delta_{mn} p_m  - 4 \sum_{m,n}  \Delta_{mn} \sqrt{p_m p_n}  + 2 \sum_{m,n}  \Delta_{mn} p_n   \notag \\
&=& - 4 \sum_{m,n}  \sqrt{p_m}   \Delta_{mn} \sqrt{p_n}   \label{eq:entropy_production4}.
\eea
where we used \eqref{eq:symmetry2}. 

\section{Auxiliary Wave Function}\label{sec:wavefunction}

The action \eqref{eq:action1} with entropy production approximated by \eqref{eq:entropy_production4} can be rewritten as
\be
{\cal S}  =   \int_0^T  dt\;\left ( \sum_{m,n} \sqrt{p_m(t)} \left (  - 4  \Delta_{mn} + \delta_{mn}  \sum_\alpha  \lambda_\alpha(t) \Theta^{(\alpha)}_n  \right )\sqrt{p_n(t)}  -  \sum_\alpha \lambda_\alpha(t) \theta_\alpha \right ) \label{eq:action2}.
\ee
To understand the behavior of the system it is useful to introduce a new set of Lagrange multipliers defined by
\be
\Lambda_\alpha(t) \equiv  \int_0^t \lambda_\alpha(\tau) d\tau + \Lambda_\alpha(0),\label{eq:new_multipliers}
\ee
and then the action \eqref{eq:action2} takes the following from
\bea
{\cal S}  =   \int_0^T  dt\;\left ( \sum_{m,n} \sqrt{p_m(t)} \left (  - 4  \Delta_{mn} +\delta_{mn}   \sum_\alpha \Theta^{(\alpha)}_n   \frac{d \Lambda_\alpha(t)  }{dt} \right )\sqrt{p_n(t)} \right)\notag\\
+  \sum_\alpha \left ( \Lambda_\alpha(0)- \Lambda_\alpha(T) \right ) \theta_\alpha  \label{eq:action3}.
\eea
Moreover, the symmetric transition matrix can be decomposed as
\be
\Delta_{mn}  = \sum_{p,q} Q^T_{mp} D_{pq} Q_{qn},
\ee
where $D$ is a real diagonal matrix, $Q$ is an orthogonal matrix, 
\be
\sum_{p} Q^T_{mp} Q_{pn} = \delta_{mn},
\ee
and $Q^T$ is the transpose of $Q$,
\be
Q^T_{mn} = Q_{nm}.
\ee
Then the action \eqref{eq:action3} becomes
\bea
{\cal S}  =   \int_0^T  dt\;\left ( \sum_{m,n,p,q} \sqrt{p_m(t)} Q^T_{mp}  \left (  - 4  D_{pq} +  \delta_{pq}  \sum_\alpha \Theta^{(\alpha)}_q   \frac{d \Lambda_\alpha(t)  }{dt} \right )  Q_{qn} \sqrt{p_n(t)} \right) \notag\\
+  \sum_\alpha \left ( \Lambda_\alpha(0)- \Lambda_\alpha(T) \right ) \theta_\alpha \label{eq:action4}.
\eea
(Note that $ \Lambda_\alpha(0)$, $\Lambda_\alpha(T)$ and $ \lambda_\alpha(t) = \frac{d \Lambda_\alpha(t)  }{dt}$ are not all independent (see Eq. \eqref{eq:new_multipliers}) and so one can only vary $ \Lambda_\alpha(t)$ for either $t \in [0,T)$, i.e. solving for a stationary path for a given initial state, or  $t \in (0,T]$, i.e. solving for a stationary path for a given final state.)

It is now convenient to define an auxiliary wave-function
\bea
\psi_m &\equiv& \sum_{n} e^{i \varphi_m} Q_{mn} \sqrt{p_n}  
\eea
with phases given by
\be
\varphi_m(t) \equiv  \frac{1}{4} \sum_\alpha \Theta^{(\alpha)}_m  \Lambda_\alpha(t)  =   \sum_\alpha \Theta^{(\alpha)}_m  \left (  \int_0^t \lambda_\alpha(\tau) d\tau + \Lambda_\alpha(0) \right ) \label{eq:phases}
\ee
and then
\bea
{\cal S}  &=& \int_0^T  dt\;\left (-4  \sum_{m,n} \left (  \psi_m^* D_{mn} \psi_n   -  \frac{d \varphi_n}{dt}  \psi_m^* \delta_{mn} \psi_n   \right )\right ) + \sum_\alpha \left ( \Lambda_\alpha(0)- \Lambda_\alpha(T) \right ) \theta_\alpha   \notag \\
&=& \int_0^T  dt\;\left (-4  \sum_{m,n} \left (  \psi_m^* D_{mn} \psi_n   + i   \psi_m^* \delta_{mn} \frac{d \psi_n}{dt} \right )  - i \sum_{m,n,l} \sqrt{p_m} Q^T_{ml} Q_{ln}  \frac{d \sqrt{p_n}}{dt} \right ) + \sum_\alpha \left ( \Lambda_\alpha(0)- \Lambda_\alpha(T) \right ) \theta_\alpha  \notag\\
&=& \int_0^T  dt\;\left (-4  \sum_{m,n} \left (  \psi_m^* D_{mn} \psi_n   + i   \psi_m^* \delta_{mn} \frac{d \psi_n}{dt} \right )  - i \sum_{m} \sqrt{p_m}  \frac{d \sqrt{p_m}}{dt}\right )  +\sum_\alpha \left ( \Lambda_\alpha(0)- \Lambda_\alpha(T) \right ) \theta_\alpha  \notag\\
&=& \int_0^T  dt\;\left (-4  \sum_{m,n} \left (  \psi_m^* D_{mn} \psi_n   + i   \psi_m^* \delta_{mn} \frac{d \psi_n}{dt} \right )  - \frac{1}{2} i \sum_{m} \frac{d p_m }{dt} \right ) +\sum_\alpha \left ( \Lambda_\alpha(0)- \Lambda_\alpha(T) \right ) \theta_\alpha   \notag\\
&=& \int_0^T  dt\;\left (-4  \sum_{m,n} \left (  \psi_m^* D_{mn} \psi_n   + i   \psi_m^* \delta_{mn} \frac{d \psi_n}{dt} \right ) \right ) +\sum_\alpha \left ( \Lambda_\alpha(0)- \Lambda_\alpha(T) \right ) \theta_\alpha  \label{eq:action5}.
\eea

\section{Schr\"odinger Equation}\label{sec:Schrodinger}

The action \eqref{eq:action5}  can be expressed using the bra-ket notations
\be
{\cal S} =   \int_0^T  dt\;\left (  -4  \left (\langle \psi| \hat{\Delta} |\psi\rangle  + i \langle \psi| \frac{d}{dt}|\psi\rangle \right ) \right ) +  \sum_\alpha \left ( \Lambda_\alpha(0)- \Lambda_\alpha(T) \right ) \theta_\alpha  \label{eq:action6}
\ee
where
\bea
|\psi \rangle &\equiv& \sum_{n=1}^N  \psi_n |n\rangle \\
\langle \psi |  &\equiv& \sum_{n=1}^N  \langle n | \psi^*_n\\
\hat{\Delta} &\equiv& \sum_{n=1}^N D_{nn} | n \rangle \langle n |
\eea
and $|n\rangle$ are eigenvectors of the transition operator $\hat{\Delta}$ with eigenvalues $D_{nn}$.  Although \eqref{eq:action6}  was derived in a particular basis, it  is invariant under $U(N)$ transformations and so the corresponding equations should also be valid in any basis.  

By setting variations of \eqref{eq:action6} with respect to $\psi_n$'s to zero (and ignoring the boundary terms) we arrive at our main result:
\be
i\frac{d}{dt} |\psi\rangle = - \hat{\Delta} |\psi\rangle. \label{eq:Schrodinger}
\ee
Note, that in the original problem the path of the stationary entropy production should have been determined by setting variations of \eqref{eq:action6} with respect to probabilities $p_n$ and Lagrange multipliers $\lambda_\alpha$ (or equivalently $\Lambda_\alpha$) to zero. However, to obtain \eqref{eq:Schrodinger} the action \eqref{eq:action6} was varied with respect to $\psi_n$ (or $p_n$ and $\varphi_n$). Strictly speaking, these two procedures are not equivalent,
\be
\frac{\delta {\cal S}}{\delta\psi_n}=0\;\;\;\;\;\Leftrightarrow\;\;\;\;\;\frac{\delta {\cal S}}{\delta p_n}=\frac{\delta {\cal S}}{\delta\varphi_n}=0\;\;\;\;\;\stackrel{?}{\Leftrightarrow}\;\;\;\;\;\frac{\delta {\cal S}}{\delta p_n}=\frac{\delta {\cal S}}{\delta\Lambda_\alpha}=0\;\;\;\;\;\Leftrightarrow\;\;\;\;\;\frac{\delta {\cal S}}{\delta p_n}=\frac{\delta {\cal S}}{\delta\lambda_\alpha}=0\notag.
\ee
From \eqref{eq:phases} we do get that 
\be
\frac{\delta {\cal S}}{\delta\varphi_n}=0 \;\;\;\;\;\Rightarrow\;\;\;\; \frac{\delta {\cal S}}{\delta\Lambda_\alpha} = \sum_m \frac{\delta {\cal S}}{\delta\varphi_m} \frac{\delta \varphi_m}{\delta\Lambda_\alpha} = \frac{1}{4} \sum_m \frac{\delta {\cal S}}{\delta\varphi_m} \Theta_m^{(\alpha)} = 0, 
\ee
but the opposite, i.e.
\be
\frac{\delta {\cal S}}{\delta\Lambda_\alpha}=0  \;\;\;\;\;\Rightarrow\;\;\;\; \frac{\delta {\cal S}}{\delta\varphi_n}=0,
\ee
is only true if $\Theta_n^{(\alpha)}$ is an invertible square matrix, or if there are $K=N$ linearly independent constraints. In our case $K<N$ and so \eqref{eq:Schrodinger} is at most an approximation which can only be a valid approximation if the number of constraints is sufficiently large $K \lesssim N$. Indeed, one use $K$ equations to fix the values of  $K$ phases, but then the remaining $N-K$ phases could be arbitrary. However, if not all of the phases are directly observable, then the fact that some phases are not uniquely determined is not a problem. For example, the overall phase is not observable and so assigning to it an arbitrary value does not produce any observable effects.

It is important to emphasize that in the above description the auxiliary  wave function $\psi_n$ was only a bookkeeping device which keeps track of the information about a system with a large number of hidden symmetries/constraints. These constraints were first described explicitly by $\Theta_n^{(\alpha)}$ and $\theta_\alpha$, but in the Schr\"odinger-like equation \eqref{eq:Schrodinger} they only appear implicitly in the form of phases $\varphi_n$ defined as linear combinations of the Lagrange multipliers $\Lambda_\alpha$.  This suggests that the state vector $|\psi\rangle$ should not be considered as an {\it ontic} state, i.e. representing the real state of a quantum system, but rather as an {\it epistemic} state, i.e. representing the state of knowledge about the state of a statistical system with hidden constraints.

In conclusion, let us mention an interesting connection between equations \eqref{eq:master} and \eqref{eq:Schrodinger}. It is well known that one can go from \eqref{eq:master} to \eqref{eq:Schrodinger} by performing the so-called Wick rotation, i.e. by replacing $t$ with $it$ in  \eqref{eq:master} we arrive at \eqref{eq:Schrodinger}. (The fact that one equation is written for $p_n$'s and the other one for $\psi_n$'s is of no importance for this argument.) But now we also have a physical interpretation of what the Wick rotation actually means. It takes us from a statistical system with no hidden constraints (e.g. Markov process described by \eqref{eq:master}) to a statistical system with many hidden constraints (e.g. non-Markov process described by \eqref{eq:Schrodinger}).

\section{Discussion}\label{sec:conclusion}

In this paper we considered a stochastic process whose dynamics is described by a Markov chain on only short time-scales and, at the same time, constrained to conserved a number of quantities on long time-scales. To study the system we applied the principle of stationary entropy production and used the method of Lagrange multipliers to derive the action \eqref{eq:action6} and the corresponding equation \eqref{eq:Schrodinger}. Despite of the apparent similarities between \eqref{eq:Schrodinger} and the usual Schr\"odinger equation, the considered statistical system is not equivalent to quantum mechanics. First of all, to derive \eqref{eq:entropy_production3} the detailed balance condition had to be assumed and thus our results would have to be modified for more general systems.  Secondly, the entropy production was approximated by  \eqref{eq:entropy_production4} which is only valid up to the fourth order in $p_n-p_m$, i.e. not too far from an equilibrium. Thirdly, equation  \eqref{eq:Schrodinger} is only accurate when the number of linearly independent constraints is sufficiently large. Therefore, we must conclude that from our statistical system we do not get quantum mechanics {\it per se}, but something which can in certain limits reduce to quantum mechanics. 

If our analysis is correct (and it is a big if) then the framework of entropic mechanics presented here may be considered as more general than quantum mechanics. Since it is based entirely on only the principle of stationary entropy production nothing stops us from applying this principle to other (than considered here) stochastic processes. In fact one such process was analyzed in Ref. \cite{Vanchurin} where it was argued that the entropy production could gives rise to an emergent dynamics of the metric. By taking a phenomenological approach (in Sec. 6 of Ref.  \cite{Vanchurin}) the dynamics was described approximately by expanding the entropy production into products of generalized forces (derivatives of metric) and conjugate fluxes. Near equilibrium these fluxes are given by an Onsager tensor contracted with generalized forces and on the grounds of time-reversal symmetry the Onsager tensor is expected to be symmetric. Then it was shown that a particularly simple and highly symmetric form of the Onsager tensor gives rise to the Einstein-Hilbert term, i.e. general relativity. 

We would like to stress that our results (here and in Ref. \cite{Vanchurin}) suggest that both quantum mechanics and general relativity may not be exact, but only approximate near-equilibrium limits of some statistical systems for which conservations/symmetries play a crucial role. As such we expect that both theories should break down further away from the equilibrium where some of the symmetries are expected to be broken. It would be interesting to see if one can attribute the breaking of these symmetries to other mysteries such as dark matter, dark energy or cosmic inflation. On the other hand, if quantum mechanics and general relativity are only limits of some other theory, then there is no need to look for a theory of quantum gravity - it simply does not exist.

{\it Acknowledgments.}  The author wishes to acknowledge the hospitality of the Pacific Science Institute where this work began, the University of Ni\v s where the key results were obtained and the Duluth Institute for Advance Study where much of the work in completing the paper was carried out. The work was supported in part by the Foundational Questions Institute (FQXi).

\end{document}